\documentstyle[12pt,aaspp4]{article}
\slugcomment{Accepted for publication in the Astrophysical Journal}

\lefthead{Sankrit et al.} \righthead{FUSE Vela XBRT}

\begin{document}

\title{FUSE Observations of an X-Ray Bright Region in the
Vela Supernova Remnant \altaffilmark{1}}

\author{Ravi Sankrit\altaffilmark{2}, 
Robin L. Shelton\altaffilmark{2}, 
William P. Blair\altaffilmark{2}, 
Kenneth R. Sembach\altaffilmark{2}
\and
Edward B. Jenkins\altaffilmark{3}}

\altaffiltext{1}{Based on data obtained for the Guaranteed Time Team by the
NASA-CNES FUSE mission operated by the Johns Hopkins University.  Financial
support to U.S. participants has been provided by NASA contract NAS5-32985.}
\altaffiltext{2}{The Johns Hopkins University, Department of Physics and Astronomy,
3400 N. Charles Street,  Baltimore, MD 21218}
\altaffiltext{3}{Princeton University Observatory, Princeton, NJ 08544}


\begin{abstract}
We present the results of a \textit{Far Ultraviolet Spectroscopic
Explorer} observation of an X-ray selected knot in the Vela supernova
remnant.  Spectra were obtained through the
30\arcsec$\times$30\arcsec\ low resolution aperture and the
4\arcsec$\times$20\arcsec\ medium resolution aperture.
O~VI~$\lambda\lambda$1032,1038 and C~III~$\lambda$977 are detected
strongly in both spectra, and S~VI~$\lambda\lambda$933,944 is detected
weakly only in the larger aperture spectrum.  We also report the first
detection of C~II~$\lambda$1037 emission in a supernova remnant.  The
spectra show the presence of two kinematic components along the line of
sight - one with both low and high excitation emission centered at a
velocity of $-50$~km~s$^{-1}$ and another with only low excitation
emission centered at a velocity of $+100$~km~s$^{-1}$.  We associate
the $-50$~km~s$^{-1}$ component with the observed X-ray knot, and find
a dynamical pressure of $3.7\times 10^{-10}$~dyne~cm$^{-2}$ driving the
shock.  We compare our results with data obtained using the
\textit{Hopkins Ultraviolet Telescope} at nearby locations and find
that differences in the spectra imply the existence of two emitting
components in the X-ray knot.  Based on the X-ray morphology seen in a
ROSAT HRI image, we identify two distinct regions which can be
associated with these two components whose ultraviolet emission differs
dramatically.  These observations demonstrate the importance of high
spectral resolution in understanding the proper physical relationships
between the various emitting components in supernova remnants.
\end{abstract}



\section{Introduction}

The Vela supernova remnant (SNR) is a nearby Galactic SNR that is
visible in all passbands from radio to X-ray.  Its age, based on the
spin down age of the central Vela pulsar, is about 11,400 years
(\cite{rei70}) and it is at a distance of about 250~pc (\cite{cha99}).
In the optical it is characterized by long arc-shaped filaments with
arbitrary centers of curvature and has been classified as having a
``smoke-ring'' morphology by \cite{van78}.  In the X-ray (\cite{kah85};
\cite{asc95}), the overall shape and extent of the remnant is more
easily discerned -- it is roughly circular with an angular diameter of
about 8\arcdeg.  However, the detailed structure of the X-ray emitting
gas is very complex, with a plerion around the central pulsar
(\cite{har85}), and several knots and filaments having a range of
temperatures between one million and a few million degrees
(\cite{kah85}).  To add to the complexity, another SNR (RXJ0852.0-4622)
lying within the boundaries of Vela in projection has been discovered
in ROSAT X-ray maps (\cite{asc98}; \cite{sla00}).  The overall extent
of the radio emission in Vela is similar to the extent of the X-ray
emission (\cite{dun96}), but when considered in detail the radio
filaments only occasionally correlate well with the X-ray emission and
the optical filaments (\cite{boc98}).  The velocity structure of the
SNR is also very complex.  Absorption line studies towards several
stars in the region have shown that high velocity components are
distributed more or less randomly across the face of the remnant
(\cite{jen76}; \cite{cha00}).

Vela, as a ``middle-aged'' remnant whose emission is dominated by the
interaction of the supernova blast wave with the surrounding
interstellar medium (ISM), has often been compared and contrasted with
the Cygnus Loop, another nearby remnant at a similar stage of
evolution.  The Cygnus Loop has a classical limb-brightened shell
morphology at radio, optical and X-ray wavelengths, and there is
evidence to show that the SN explosion happened in a cavity cleared out
by the progenitor star (\cite{lev98}).  Vela, in contrast, seems to be
the result of an explosion in a highly inhomogeneous medium.  However,
the optical and ultraviolet emission from individual filaments in both
these remnants is due to shock excitation of the ambient medium
(e.g.~\cite{ray81}).  The source of the energy for these shocks is the
supernova explosion, and therefore, understanding the properties of
individual filaments is an important part of addressing the broader
problem of SNR evolution in the ISM.

Ultraviolet spectra of filaments in Vela have revealed the presence of
lines from species covering a wide range of ionization states, from
C~II to O~VI (\cite{ray81}; \cite{bla95}; \cite{ray97}).  These data
have been used to infer that the emission is due to shocks with
velocities between about 100~km~s$^{-1}$ and 200~km~s$^{-1}$.  In this
paper we present far-ultraviolet spectra of an X-ray knot in Vela
obtained with the \textit{Far Ultraviolet Spectroscopic Explorer}
(FUSE).  The X-ray knot is near the projected center of Vela and it has
bright optical filaments running along part of its eastern edge.
Spectra of the knot obtained with the \textit{Hopkins Ultraviolet
Telescope} (HUT) were presented by R97.  HUT apertures were placed on
the bright optical filament and a region to the west within the X-ray
knot in an attempt to view edge-on and face-on portions of the same
shock.  We have observed a region adjacent to the HUT ``face-on''
position selected to have less optical emission and more uniformly
bright X-ray emission.  The two datasets can therefore be compared;
they are complementary - HUT spectra do not have the high spectral
resolution of the FUSE spectra, but they span a wider wavelength region
and include many more diagnostic emission lines.  We also present
archival ROSAT images of the X-ray knot to guide our understanding of
the local morphology and the relationship between the X-ray and the
ultraviolet emission.  The aperture positions for both HUT and FUSE
observations are shown in Figure~\ref{figxray}.  From left to right
(east to west), these are (and we will use the following nomenclature
in this paper) - HUT ``edge-on'', HUT ``face-on'', FUSE low resolution
(LWRS) and FUSE medium resolution (MDRS) apertures.  (The location of
the optical filaments can be seen in Figure~1 of R97.)


\section{Observations}

The FUSE observations (ID P1141202) were obtained on 25 January 2000 as
part of the Guaranteed Time Team project on SNRs.  Five exposures with
a total integration time of 11023~s were obtained with the low
resolution (LWRS) 30\arcsec$\times$30\arcsec\ aperture centered at
RA(J2000)~=~$08^h\,41^m\,02\fs43$, DEC(J2000)~=~-44\arcdeg\,44\arcmin\,01\farcs8.
Data were obtained simultaneously through the medium resolution (MDRS)
20\arcsec$\times$4\arcsec\ aperture, which was located about
3.5\arcmin\ westward from the LWRS aperture (Figure~\ref{figxray}d).
The FUSE instrument and its performance have been described in detail
by \cite{moo00} and \cite{sah00}.  Briefly, there are four independent
channels -- SiC1, SiC2, LiF1 and LiF2, each having two segments --
``A'' and ``B''.  Therefore each observation may be considered as
having eight separate spectral segments.  We will refer to these
segments by names -- ``SiC1A'' and so on.  The overall wavelength
coverage is 905\AA\ - 1187\AA\@.  The shorter wavelengths
($\sim$~905\AA\ - 1000\AA) are covered by segments SiC1B and SiC2A, the
intermediate wavelengths ($\sim$~1000\AA\ - 1100\AA) by segments LiF1A,
SiC1A, LiF2B, SiC2B and the longer wavelengths ($\sim$~1100\AA\ -
1187\AA) by segments LiF1B and LiF2A\@.  We will discuss spectra taken
through the LWRS and MDRS apertures.  Since the source is extended and
thus fills the apertures, the effective spectral resolution is
$\sim$0.34\AA\ for the LWRS data and $\sim$0.045\AA\ for the MDRS
data.

To reduce the data, we first extracted one-dimensional spectra from the
raw data for each segment in each of the five exposures.  The five
exposures were then added segment by segment.  The count rate is simply
the total counts divided by the total exposure time (11023~s).  The
flux calibration was done using the standard files used by the CALFUSE
calibration pipeline software (version 1.6.9), which gives the count
rate to flux conversion as a function of pixel number for each segment
and aperture.  We used the same wavelength solutions as the calibration
pipeline, but found we needed to apply a zero point offset for each
segment.  The offsets were determined using the airglow lines -
Ly$\beta$ for LiF1A, SiC1A, LiF2B and SiC2B and Ly$\gamma$ for SiC1B
and SiC2A\@.  (Only these segments will be considered in this paper,
since no lines were detected in the long wavelength segments - LiF1B
and LiF2A\@.)  Finally all the segments contributing to a wavelength
region were co-added, weighted appropriately by their effective areas.
(In the case of the short wavelength MDRS spectrum, however, only the
SiC1A data were used.) Thus our final data products consist of LWRS and
MDRS spectra in the 905\AA~-~1000\AA\ and 1000\AA~-~1100\AA\ bands.

Images of the X-ray knot (on which the FUSE apertures are located) were
obtained by ROSAT with the PSPC (sequence ID rp500013n00) in 1991 and
with the HRI (sequence ID rh500135n00) in 1992.  Both PSPC and HRI
pointings were centered at RA(J2000)~=~$08^h\,40^m\,45\fs0$,
DEC(J2000)~=~-44\arcdeg\,38\arcmin\,24\farcs0, placing the X-ray knot
in the center of the field.  These data are now available to the public
and we have obtained them from the archive maintained by
HEASARC\footnote{High Energy Astrophysics Space Archive Research
Center}.  In the case of the PSPC images, we obtained the raw data from
each energy channel and processed them according to the method
described by \cite{sno98}.  For the HRI data, we obtained the
5\arcsec\ resolution image directly from the archive.  These X-ray
images are shown in Figure~\ref{figxray}.  The PSPC images in the R1R2
band (0.11~keV~-~0.28~keV) and the R4R5 band (0.44~keV~-~1.21~keV) are
shown in Figures~\ref{figxray}a and \ref{figxray}b respectively.  The
field of view for the PSPC images is 2\arcdeg.  The HRI image of the
knot is shown in Figure~\ref{figxray}c, and a blowup of the HRI image,
with overlaid contours and the location of the HUT and FUSE apertures
is shown in Figure~\ref{figxray}d.  We discuss the X-ray morphology in
more detail in \S4.


\section{Results}

The lines detected in the FUSE LWRS spectrum are
S~VI~$\lambda\lambda$933,944, C~III~$\lambda$977 and
O~VI~$\lambda\lambda$1032,1038.  The spectral regions near these lines
are shown in Figure~\ref{figlwrs}.  The O~VI lines are very strong and
each line has a FWHM of about 0.45\AA\ ($\sim$130~km~s$^{-1}$). These
lines are broader than the terrestrial airglow lines (e.g. Ly$\beta$,
Ly$\gamma$), which have filled slit widths of about 0.34\AA.  The C~III
line is double peaked.  The shorter wavelength component is about the
width of the airglow line, and the longer wavelength component is
broader, having a FWHM of about 150~km~s$^{-1}$.  The component
centroids are separated by about 0.5\AA.  The S~VI lines are very
weak and the widths cannot be measured as accurately.  In
Figure~\ref{figmdrs} the regions around the C~III and O~VI lines from
the MDRS spectra are shown (S~VI is too weak to be detected).  The O~VI
lines, as in the LWRS spectrum, are broader than the airglow.  Each has
a FWHM of about 0.2\AA\ ($\sim$60~km~s$^{-1}$), compared to about
0.045\AA\ for the filled slit width.  The O~VI line widths imply that
the gas temperature, T~$\lesssim~1.2\times10^{6}$~K.  If the
properties of the gas sampled by the LWRS and MDRS spectra are the
same, then the width of the O~VI lines in the LWRS spectrum is the
result of the instrument line profile convolved with their intrinsic
width, which is at most about 0.2\AA.  The C~III emission in the MDRS
is also double peaked, with the two components separated by the same
amount as in the LWRS spectrum.  As in the LWRS spectrum, the shorter
wavelength component is about the width of the slit, and the longer
wavelength component is broader, consistent with the FWHM of about
150~km~s$^{-1}$ seen in the LWRS spectrum.

In Figure~\ref{figo6o6} we show an overlay of O~VI~$\lambda$1032 flux
and 2 times the O~VI~$\lambda$1038 flux plotted as a function of
velocity.  This is shown for the LWRS (top panel) and MDRS (bottom
panel) spectra.  The lower state for each of these transitions is the
ground state, and the upper states have statistical weights in a 2:1
ratio.  Therefore, in the case of optically thin emission, the
1032\AA\ line is expected to be twice as strong as the 1038\AA\ line.
As the O~VI optical depth increases, the shorter wavelength line is
preferentially scattered out of the line of sight and this ratio
becomes smaller.  The overlay in Figure~\ref{figo6o6} shows that for
the region we have observed, the emission is close to being optically
thin.  The lines are centered at about $-50$~km~s$^{-1}$ with FWHM of
about 130~km~s$^{-1}$ in the LWRS and 60~km~s$^{-1}$ in the MDRS.

The line profiles of the two O~VI lines match up well at velocities
around the emission peak and in the red wing, but the
O~VI~1038\AA\ line has excess emission on its blue wing, between
$-200$~km~s$^{-1}$ and $-300$~km~s$^{-1}$ (Figure~\ref{figo6o6}, top
panel).  This excess is seen clearly in each of the individual LWRS
segments covering the O~VI wavelength region and it is therefore highly
unlikely to be an instrumental artifact.  We have identified this
feature as C~II~$\lambda$1037.02 emission.  It is at the correct
wavelength for this line blue shifted by the same amount as the O~VI
lines.  A corresponding excess of emission at the same velocities is
seen also in the MDRS spectrum.  (By itself, the MDRS spectrum is noisy
as seen in Figure~\ref{figmdrs} and we would not have claimed a
detection based on that alone.  However, the coincidence with the LWRS
result makes the identification plausible in the MDRS data.) The
C~II~1037.02\AA\ line is one of a pair of lines that have a common
upper state.  This line has a lower state which is 63~cm$^{-1}$ above
the ground state, while the companion 1036.34\AA\ line is a ground
state transition (see \cite{mor91}).  We do not see the
1036.34\AA\ line in our spectra, but this is not surprising.  This line
is only half as strong as the 1037.02\AA\ line in the recombination
spectrum and furthermore, being a ground state transition, it could
quite easily be absorbed by intervening gas.  The existence of C~II is
expected in a fully recombined shock.  For instance, R97 detected
C~II~$\lambda$1335 (albeit weakly) in their HUT spectra of nearby
regions in Vela.

In Figure~\ref{figo6c3} we show an overlay of C~III~$\lambda$977 and
O~VI~$\lambda$1032 fluxes plotted in velocity space.  The C~III flux
has been multiplied by 2 for the display.  The two components of C~III
are seen in both the LWRS (top panel) and MDRS (bottom panel) spectra.
The shorter wavelength component of C~III lines up with the O~VI line,
though it is clearly narrower.  (The offset of about 20~km~s$^{-1}$
between their peaks is not significant.  The lines are from two
different detector segments, and the offset is within the relative
errors in the absolute wavelength scale.)  The second component of
C~III is wider (FWHM~$\sim$~150~km~s$^{-1}$) and is centered at about
+100~km~s$^{-1}$.  We estimate that this red-shifted component contains
about 45\% of the total C~III flux.

C~III~$\lambda$977 is a strong resonance line and therefore the
observed line profile is likely to be affected by self absorption (see
\cite{bla00a} for an example of this effect in the LMC SNR N49).
Having observed no other low ionization lines in the FUSE spectrum we
cannot make a detailed assessment of the effect of self absorption.
However, the fact that the width of the red-shifted component is the
same in both LWRS and MDRS spectra while the blue-shifted component is
narrower in the MDRS (reflecting filled slit emission) is evidence that
the two emission peaks are from two physically distinct components.
(This interpretation is also supported by the O~VI line profiles which
show no sign of a tail towards the red-shifted wing).  We have examined
FUSE spectra of a few stars behind Vela and found that the C~III
absorption is typically centered near zero velocity, has a FWHM of
about 60~km~s$^{-1}$ and a peak absorption of about 50\%.  Absorption
by such a component would change the C~III flux by about 8\%.  Also,
taking this absorption into account would leave unchanged our
conclusion that there are two distinct components.  We note that the
C~III fluxes presented in this paper have not been corrected for any
self absorption.

In Table~\ref{tblflux} we list line fluxes and the C~III and O~VI
surface brightnesses for our FUSE spectra and for the HUT spectra of
adjacent regions presented in R97.  The FUSE line fluxes were obtained
by simple trapezoidal integration over the lines.  The error in each of
the fluxes for the stronger lines is dominated by the absolute flux
calibration which is accurate to about 10\% (\cite{sah00}).  For the
MDRS C~III and LWRS S~VI fluxes, which have higher random errors, we
estimate the accuracy to be 15\% and 30\% respectively.  In the table,
the total C~III flux is presented for the FUSE observations, to allow
direct comparison with the HUT data, for which the kinematic components
were unresolved.  (In the FUSE spectra, the blue-shifted component
contains $\sim$~55\% of the flux, and the red-shifted component
contains the remaining $\sim$~45\%).  Also, the C~II~$\lambda$1037.02
flux, estimated to be $7\times10^{-15}$~erg~s$^{-1}$~cm$^{-2}$ in the
LWRS spectrum (Figure~\ref{figo6o6}), was subtracted from the
O~VI~$\lambda$1038 flux.  (The exact value of the C~II flux will not
affect our discussion below, since it is so much weaker than the O~VI.)

The flux ratio of O~VI~$\lambda$1032 to O~VI~$\lambda$1038 in the FUSE
spectra (LWRS and MDRS) is about 2:1, indicating that the emission is
optically thin.  (We note that the two lines were not resolved with HUT
and so only the total flux was presented.)  The S~VI~$\lambda$933 and
S~VI~$\lambda$944 line strengths are also consistent with a 2:1 ratio
as expected for optically thin emission.  The O~VI surface brightnesses
in the LWRS and MDRS spectra differ by less than 5\%.  However, the
C~III surface brightness in the MDRS spectrum is about 30\% higher than
in the LWRS spectrum.  It is not clear how significant this difference
is - the MDRS short wavelength spectrum is noisy and the C~III line is
weak, so its measured flux is sensitive to the background value chosen
and the real difference may not be as high as 30\%.  In any case,
because the similarity in the line profiles, we will assume that the
LWRS and MDRS apertures sample gas having more or less the same
emission properties and use mainly the LWRS data below.

The C~III to O~VI surface brightness ratio is 0.24 for the FUSE LWRS
spectrum, 0.10 for the HUT face-on shock position and 0.39 for the HUT
edge-on shock position.  Differences in the C~III surface brightness
contribute more to this range of values than do differences in the O~VI surface
brightness.  The C~III in the FUSE LWRS spectrum is 3.1 times as bright
as in the HUT face-on spectrum, while the O~VI is only 1.3 times as
bright.  Similarly, the C~III in the HUT edge-on spectrum is 5.5 times
as bright as in the FUSE LWRS spectrum, while the O~VI is 3.4 times as
bright.  The probable cause of these differences is discussed below in
\S4.2.

The fluxes presented in Table~\ref{tblflux} have not been corrected for
interstellar reddening.  While this is useful in comparing different
observations, the correction needs to be done to obtain intrinsic line
fluxes that reflect conditions in the emitting gas.  Even the moderate
color excess, E$_{B-V} = 0.1$, towards Vela (\cite{wal90},
\cite{bla00b}) results in significant extinction at far-ultraviolet
wavelengths.  R97 corrected their HUT spectra of Vela using the
\cite{car89} extinction curve and, following previous studies of the
Cygnus Loop SNR (\cite{bla91}, \cite{lon92}), used the \cite{lon89}
extinction curve at the shortest wavelengths.  They obtained correction
factors of 3.8 and 3.7 for the C~III and O~VI fluxes, respectively.
The \cite{lon89} extinction curve, which is based on \textit{Voyager}
observations, flattens out at about 1000\AA\ and falls below the
extrapolation of the \cite{sav79} extinction curve (see their
Figure~3).  As \cite{lon89} themselves discuss, it is unclear which of
these extinction curves is more accurate, especially since the data
used to derive them had very low spectral resolution.  For this paper,
we use the more recent extinction curve presented by \cite{fit99}, and
follow his suggestion that extrapolation to far-ultraviolet wavelengths
is the best dereddening strategy.  For E$_{B-V} = 0.1$ and total to
selective visual extinction, $R = 3.1$ (the standard value for diffuse
ISM gas), we obtain correction factors of 4.5 and 3.7 for the C~III and
O~VI fluxes, respectively.  (The calculations were done using the IDL
Astronomy Library implementation of the Fitzpatrick reddening
correction routine.)


\section{Discussion}

\subsection{Shock Components Observed by FUSE}

The FUSE data show the existence of two components contributing to the
ultraviolet emission along the line of sight.  One component, blue
shifted by about 50~km~s$^{-1}$ has strong O~VI and C~III emission.
Weaker S~VI and C~II emission is also detected from this component.
The C~III emission can be explained by a shock that has cooled and
recombined.  Assuming that 55\% of the total C~III emission
(Table~\ref{tblflux}) is from this component and correcting for
reddening using the factors given above, the intrinsic O~VI to C~III
flux ratio of this component is 6.3, implying a shock velocity of about
180~km~s$^{-1}$.  The shock front would have to be moving
$\sim$15\arcdeg\ out of the plane of the sky, towards us, for the
emission peak to be at $-50$~km~s$^{-1}$.  The presence of S~VI
emission is consistent with a shock velocity of 180~km~s$^{-1}$ as it
is a comparable but slightly lower ionization species than O~VI.  The
second component, red-shifted by about 100~km~s$^{-1}$, has strong
C~III emission but no O~VI emission (Figure~\ref{figo6c3}).  The
emission is thus due to a shock with velocity less than
about 140~km~s$^{-1}$, which produces negligible O~VI.  The location of
this shock along the line of sight is uncertain - it could, for
instance, be a shock driven into a cloud on the back face of the
remnant.  It is worth noting that \textit{Voyager} UVS spectra of Vela
presented by \cite{bla95} showed C~III to O~VI ratios ranging from 0.6
to 2.8 in 300 square arcminute fields of view, indicating that the
production of ultraviolet lines was dominated by slower shocks.

For the component with an observed Doppler shift
$v_{obs}=-50$~km~s$^{-1}$, we can invoke a procedure described by
\cite{ray88} to determine the shock's dynamical pressure from an
emission line, regardless of the projection factor $\cos\theta$ of the
front's normal vector onto the line of sight.  (See also equations 1-3
of R97.) Calculations of emission line intensities from planar shocks
(\cite{har87}) indicate that for shocks with velocities over the range
$160 < v_s < 400$~km~s$^{-1}$, the production rate $Y$ for C~III
$\lambda 977$ radiation is nearly constant at 0.28 photons emitted in
$4\pi$~steradians for each atom that passes through the shock front.
The yield is much higher for shocks with $v_s < 160$~km~s$^{-1}$, but
the presence of O~VI emission indicates that $v_s$ is above this
value.  The production rate assumes a carbon abundance of 8.52 on a
logarithmic scale where the abundance of hydrogen equals 12.0.  It also
assumes that the shock is radiative, the C~III zone is complete and the
emission is optically thin.  The observed C~III flux relative to the
O~VI flux requires that the shock we are considering is radiative.  In
a non-radiative shock that is fast enough to produce O~VI, the C~III
flux would be over an order of magnitude weaker.  We have run shock
models (using an updated version of the code described in
\cite{ray79}), and found that at the point the O~VI flux reaches half
its maximum value for a radiative shock, the C~III flux is less than
2\% of its maximum value.  The measured flux ratio of
C~II~$\lambda$1037 to C~III~$\lambda$977 is about 0.1.  We find from
our shock models that such a high ratio is possible only in the case
when the C~III zone is complete and most of the carbon has recombined
to C~II\@.  Since the blue-shifted component is radiative and complete,
we are justified in using the value of $Y$ from \cite{har87}, given
above.

From Table~\ref{tblflux}, we find that the average of the two FUSE
surface brightness measurements of this blue-shifted component (which
accounts for 55\% of the total observed C~III emission) is
$0.83\times10^{-16}$~erg~s$^{-1}$~cm$^{-2}$~arcsec$^{-2}$.  The energy
of a C~III~$\lambda$977 photon is $2.03\times 10^{-11}$~ergs and 1
steradian equals $4.26\times 10^{10}$~square arcseconds.  Therefore,
the blue-shifted C~III surface brightness corresponds to a specific
intensity, I$_{obs} = 1.7\times
10^5$~photons~cm$^{-2}$~s$^{-1}$~str$^{-1}$.  If we correct this
observed intensity for extinction (multiplying by 4.5) and then divide
by $Y/4\pi$, we obtain a measure of $n_0 v_s/\cos\theta$, where $n_0$
is the number density of atoms and ions entering the shock on the
upstream side.  Multiplying this quantity by the mean atomic mass
$1.3m_{\rm H}$ and $v_{obs} = v_s\cos\theta$ gives the dynamical
pressure $\rho v_s^2 = 3.7\times 10^{-10}$~dyne~cm$^{-2}$, or $p/k_{B}
= 2.7\times 10^{6}$~cm$^{-3}$~K (where k$_{B}$ is the Boltzmann
constant).  For a shock velocity of 180~km~s$^{-1}$, this implies a
pre-shock hydrogen number density of $\sim0.5$~cm$^{-3}$.

Our value for the dynamical pressure is more than a factor of 4 smaller
than the $1.6\times 10^{-9}$~dyne~cm$^{-2}$ ($p/k_{B} = 1.2\times
10^{7}$~cm$^{-3}$~K) found by R97 for a region immediately behind the
HUT edge-on shock position.  It is a factor of 8 smaller than the $2 -
4\times10^{-9}$~dyne~cm$^{-2}$ ($p/k_{B} = 1.4 - 2.8\times
10^{7}$~cm$^{-3}$~K) found by \cite{jen95} in another location within
the Vela remnant.  These other values were found using the method
described above, applied to [O~III]~$\lambda\lambda$4959,5007
emission.  In sharp contrast, our value for the dynamical pressure is
larger, by a factor of about 4, than the pressure for the diffuse
emission in Vela found by \cite{kah85}.  The pressure scales with the
diameter of the remnant and Kahn assumed a distance of 500~pc to Vela.
For the revised distance of 250~pc, his results imply a pressure of
$\sim1\times10^{-10}$~dyne~cm$^{-2}$ ($p/k_{B}~\sim~ 7\times
10^{5}$~cm$^{-3}$~K) in the bulk of the remnant.

These differences in pressure have important implications for the
properties of Vela, and for models of SNRs.  For example the assumption
sometimes made that the interior of an SNR is isobaric is not tenable
in this case.  The most straightforward explanation for a region of
significantly higher pressure is the existence of a reverse shock
driven back into the SNR interior when the forward blast wave
encounters a dense cloud (e.g.~\cite{spi82}).  The region between the
forward and reverse shocks is expected to be overpressured relative to
the interior of the remnant.  Since the pressures driving the shocks
observed by FUSE and HUT are several times higher than the pressure in
the interior, we infer the existence of reverse shocks.  The density is
also higher in this doubly shocked region which results in the X-ray
emission being enhanced.  We conclude that the X-ray knot consists of
regions between forward and reverse shocks.

\cite{cra94} studied the X-ray emission from this knot as well as two
others in Vela and found that the X-ray brightness of the knots was
about 20 times higher than neighbouring regions.  Based on these
measurements, he also came to the conclusion that these knots were
overpressured compared to other locations in the remnant due to reverse
shocks.  In this scenario, the correlation between the X-ray emission
and the radiative shocks (seen in HUT and FUSE ultraviolet spectra and
in the case of the HUT edge-on shock also as optical filaments) favours
a model where, in this region, the blast wave has encountered a cloud
but has not yet engulfed it (\cite{gra95}).  The alternative model that
the emission is from thermally evaporating material from smaller clouds
engulfed by the blast wave cannot explain the observed pressure
contrasts (\cite{hes86}; \cite{gra95}).


\subsection{Structure in the X-ray Knot}

The difference in the C~III to O~VI ratio among the FUSE LWRS
spectrum, the HUT face-on spectrum and the HUT edge-on spectrum
is empirical evidence that the properties of the emitting gas
vary on arcminute scales in the plane of the sky.  We use the
ROSAT images of the X-ray knot (Figure~\ref{figxray}) to correlate
the ultraviolet emission with the X-ray morphology of the region.

The X-ray knot has dimensions of about
30\arcmin~$\times$~12\arcmin\ and lies close to the projected center of
Vela.  Spectral analysis of \textit{Einstein} data have constrained the
temperature of this knot to lie between about $1\times10^{6}$~K and
$2\times10^{6}$~K (\cite{kah85}).  The ROSAT PSPC images in the R1R2
and R4R5 bands (Figures~\ref{figxray}a, b) show that the emission
within the knot is patchy on scales of a few arcminutes.  On this
spatial scale, the ratio of the R4R5 emission to the R1R2 emission is
approximately constant over the knot.  We have chosen specific
arcminute-size regions within the knot and fitted the spectra obtained
from individual ROSAT PSPC bands with Raymond-Smith models
(\cite{ray77}).  We find that the emission from each of these regions
can be fitted with a temperature T~$\sim~1.2\times 10^6$~K and a
hydrogen column density $N_{H}\sim~4\times 10^{20}$~cm.  (The
equilibrium calculation is presented to show the consistency between
the ROSAT data and the \textit{Einstein} data presented by
\cite{kah85}.  Further analysis of the x-ray emission, such as
nonequilibrium ionization modelling, is beyond the scope of this
paper.) The ROSAT HRI image (Figure~\ref{figxray}c) shows that the
eastern edge of the southern part of the knot is very well defined.
Figure~\ref{figxray}d is a blow-up of the HRI image, overlaid with
contours derived from the image convolved with a 2D gaussian having a
FWHM of 3 pixels (15\arcsec).  The contour labels are count rates
proportional to the X-ray flux.  The contours clearly define the
eastern edge, and show that the HUT edge-on aperture, which includes a
bright [O~III] filament (Figure~1 of R97), lies right at the boundary
of the knot.  We note that this morphological relationship of the X-ray
and optical emission seen at a spatial resolution of 5\arcsec\ also
favours the large cloud model over the thermal evaporation model
(\cite{hes86}) discussed in \S4.1, above.

The X-ray knot shows significant substructure.  Of particular
interest to the current study is the presence of a fainter band
separating two bright regions in Figure~\ref{figxray}d, demarcated by
the level 8 contours.  The contours have been chosen to highlight the
morphology seen - the feature is clear in the higher contrast image,
Figure~\ref{figxray}c, within the white box.  There is also a hint that
the feature shows up in the PSPC images, especially the R1R2 band
(Figure~\ref{figxray}a).  The FUSE LWRS aperture lies within the
brighter region to the north of this band, while the HUT face-on
aperture crosses into the fainter band.  We will call these the
``bright'' and ``faint'' regions (based on the ROSAT HRI image) in the
rest of the paper and discuss the spectral data based on this spatial
separation.

If we define the level 8 contour to be the boundary between the bright
and faint regions, then approximately one quarter of the area of the
HUT face-on aperture (931 square arcseconds) lies on the bright region
and the remaining three quarters (2793 square arcseconds) on the faint
region (Figure~\ref{figxray}d).  We assume that the surface brightness
of the lines in the upper quarter of the aperture is the same as in the
adjacent FUSE LWRS aperture.  (Note that we are assuming the same two
kinematic components along the line of sight and are not concerned with
separating the contributions of each.) Using these FUSE LWRS values for
the C~III and O~VI surface brightness given in Table~\ref{tblflux} and
multiplying by 931~arcsec$^{2}$ (the area of the HUT face-on aperture
lying on the bright region), we find the C~III and O~VI fluxes from
this bright region within the HUT face-on aperture to be
$1.19\times10^{-13}$~erg~s$^{-1}$~cm$^{-2}$ and
$4.99\times10^{-13}$~erg~s$^{-1}$~cm$^{-2}$, respectively.  We subtract
these fluxes from the total HUT face-on spectrum fluxes
(Table~\ref{tblflux}) and then divide by 2793~arcsec$^{2}$ (the area of
the aperture lying on the faint region) to obtain a C~III surface
brightness of $0.13\times10^{-16}$~erg~s$^{-1}$~cm$^{-2}$~arcsec$^{-2}$
and an O~VI surface brightness of
$3.95\times10^{-16}$~erg~s$^{-1}$~cm$^{-2}$~arcsec$^{-2}$ for the faint
region.  The C~III surface brightness of the faint region is thus only
about 10\% that of the bright region.  The O~VI emission is relatively
uniform, as the surface brightness of the faint region is 75\% that of
the bright region.  The C~III to O~VI ratio for the bright region is 0.24
(by definition equal to the ratio in the FUSE LWRS spectrum) and for
the faint region the ratio is 0.03.

Based on their analysis of the HUT data, R97 concluded that the shock
conditions in the edge-on shock and face-on shock are quite similar.
They pointed out that the observed C~III in the face-on shock spectrum
is weaker than predicted by the best fit models; however, they did not
address the question of why the C~III to O~VI ratio in the two
locations is different.  The picture we have presented requires a more
extreme difference in this ratio between the edge-on shock and the
faint region sampled by the HUT face-on aperture.  A straightforward
explanation of the very weak C~III emission in the faint region is
that the emission is dominated by ``incomplete'' radiative
shocks - where the post shock gas has not yet fully recombined to
C~III\@.  Note that this is consistent with the absence of [O~III]
emission from this faint region (Figure~1 of R97).  This means that for
a shock velocity of about 180~km~s$^{-1}$, the swept up column is less
than $1.5\times10^{18}$~cm$^{-2}$.  If this is the case, conditions in
the edge-on shock region and the faint region need not be drastically
different.  A small contrast in pre-shock density could result in
different degrees of completeness in the recombination zone.  The
edge-on shock position was chosen to be on the bright [O~III]
filaments, and since C~III and O~III trace each other in recombining
post shock gas, the strong C~III emission in that position is a
selection effect.  A systematic study of the spatial distribution of
both high and low ionization emission is required to address this issue
in detail.


\section{Concluding Remarks}

We have presented FUSE spectra of an X-ray knot in the Vela SNR\@.
Spectra taken through the LWRS and MDRS apertures show strong O~VI and
C~III emission.  The two lines of the O~VI doublet are well separated
from each other and from the Ly$\beta$ airglow in these spectra.  We
have detected C~II~$\lambda$1037.02 emission as an excess of flux on
the red wing of the O~VI~1038\AA\ line.  This is the first detection of
this emission line in an SNR\@.  The high spectral resolution also allows
us to examine the kinematic structure of the emitting gas in much more
detail than has so far been possible.  We detect two kinematic
components, one of which has a central velocity of -50~km~s$^{-1}$ and
a shock velocity of about 180~km~s$^{-1}$ (strong O~VI emission) and
the other which has a central velocity of about $+100$~km~s$^{-1}$ and
a shock velocity $<140$~km~s$^{-1}$ (no O~VI emission).  We identify
the former with the observed X-ray knot and the latter with a separate
component, possibly the back side of the SNR shell.  The properties of
the emitting gas (including the two component structure) are very
similar in both the LWRS and MDRS aperture locations, which are
separated by a few arcminutes.

We have obtained the dynamic pressure driving the shock responsible for
the blue-shifted component observed in the FUSE spectra and found it to
be a factor of about 4.5 smaller than the pressure found by R97 in an
adjacent region near the bright optical filaments tracing the edge-on
shock.  The pressure we find is a factor of about 4 larger than the
pressure in the regions of diffuse X-ray emission within Vela found by
\cite{kah85}.  We suggest that the presence of reverse shocks create
these localized regions of high pressure within the remnant.  These
regions are associated with bright X-ray emission and with the observed
radiative shocks and suggest that the emitting regions are part of a
large cloud that the supernova blast wave has encountered relatively
recently.

We have compared the FUSE spectra with HUT spectra taken at nearby
locations and found that the emission characteristics, in particular
the ratio of low excitation to high excitation lines, change on
arcminute scales within the X-ray knot.  We have presented ROSAT images
and discussed a possible relationship between the X-ray morphology and
the ultraviolet spectra.  Specifically, we suggest that there are two
distinct regions within the X-ray knot, separated by a rather sharp
boundary, running approximately east-west.  In this picture, the FUSE
apertures are completely contained in the northern X-ray bright region
while the large HUT aperture on the ``face-on'' shock position cuts
across the boundary, sampling emission from both regions.  We infer
C~III to O~VI ratios for the X-ray bright and faint regions and suggest
that variations in shock completeness can account for the observations.
This is also consistent with the recently shocked cloud scenario.


\acknowledgements

We thank the referee for several useful suggestions.  We also thank all
the people who worked on the development of FUSE, and those who are now
operating the satellite.  We acknowledge the financial support provided
by NASA contract NAS5-32985.  This research has made use of data
obtained from the High Energy Astrophysics Science Archive Research
Center (HEASARC), provided by NASA's Goddard Space Flight Center.


\clearpage

\figcaption{X-ray images of the knot in Vela.  North is up and east is
to the left in all these images.  The coordinate labels are right
ascension and declination in J2000 coordinates.  (a) ROSAT PSPC image
in the R1R2 (0.11~keV - 0.28~keV) band.  The circular field of view is
about 2\arcdeg\ in diameter.  (b) ROSAT PSPC image of the same region
as (a) in the R4R5 (0.44~keV - 1.21~keV) band (c) The ROSAT HRI image
of the knot.  The white box corresponds to the region shown in panel
(d).  (d) A blow-up of the HRI image, 500\arcsec\ on the side,
displayed with a logarithmic stretch overlaid with contours.  The
contours were derived from the same HRI image, convolved with a
gaussian with FWHM of 3 pixels.  The contour levels are 3, 5 and 8 HRI
counts per 5\arcsec$\times$5\arcsec pixel in a 61~ks exposure.  The
black boxes show aperture locations; from east to west (left to right),
they correspond to the HUT edge-on, HUT face-on, FUSE LWRS and FUSE
MDRS positions.  Note that the MDRS aperture lies due west of the LWRS
aperture.  Arrowheads at the top of the plot are placed to help locate
these boxes on the image.
            \label{figxray}}

\figcaption{Spectra taken through the 30\arcsec$\times$30\arcsec\ LWRS
aperture showing the detected lines.  The short wavelength spectra
including S~VI emission (top panel) and C~III emission (middle panel)
are the sum of two channels and the long wavelength spectrum including
O~VI emission (bottom panel) is the sum of four channels.  In all
plots, the spectra have been binned by 4 pixels.  Airglow lines (all
from the Lyman series) have been marked.
            \label{figlwrs}}

\figcaption{Spectra taken through the 4\arcsec$\times$20\arcsec\ MDRS
aperture showing the detected lines.  (S~VI is very faint, and is not
seen in the MDRS spectrum).  The short wavelength spectrum is a single
channel spectrum binned by 8 pixels.  The long wavelength spectrum is
the sum of four channels and binned by 4 pixels.  Airglow lines have
been marked.  The center of this aperture lies about 3.5\arcmin\ west
of the LWRS aperture center, well within the X-ray knot
(Figure~\protect\ref{figxray}).
            \label{figmdrs}}

\figcaption{Overlay of the O~VI $\lambda$1038 and $\lambda$1032 lines.
Both lines are centered at about $-50$~km~s$^{-1}$ and have FWHM equal to
about 115~km~s$^{-1}$.  The flux in the 1038\AA\ line profile has an
excess between $-300$ and $-200$~km~s$^{-1}$.  This is clearly seen in the
LWRS spectrum (top panel) and, though noisy, in the MDRS
spectrum as well (bottom panel).  The spectra shown here are each the
sum of 4 channels; since the excess is seen in each of the individual
channels, it is highly unlikely to be instrumental.  We identify the
feature as C~II~$\lambda$1037 emission.
            \label{figo6o6}}

\figcaption{Overlay of C~III $\lambda$977 and O~VI $\lambda$1032 for
the LWRS (top panel) and MDRS (bottom panel) data.  The C~III line has
two components.  One component centered at about $-50$~km~s$^{-1}$ is
associated with the O~VI emission, while the other component is
centered at about $+100$~km~s$^{-1}$ and has no corresponding O~VI
emission.  The red shifted component of C~III contains $\sim$45\%
of the total C~III flux.
            \label{figo6c3}}

\clearpage

\begin{deluxetable}{ccccc}
\tablecaption{Observed Line Strengths in FUSE and HUT Spectra
                                   \label{tblflux}}
\tablewidth{0pt}
\tablehead{
  \colhead{} & \colhead{FUSE LWRS} & \colhead{FUSE MDRS}
& \colhead{HUT Face-on\tablenotemark{a}} & \colhead{HUT Edge-on\tablenotemark{a}}
}
\startdata
Aperture & $30\arcsec\times30\arcsec$ & $4\arcsec\times20\arcsec$
           & $19\arcsec\times196\arcsec$ & $10\arcsec\times56\arcsec$ \\
\cutinhead{Flux (10$^{-13}$ erg s$^{-1}$ cm$^{-2}$)}
S VI 933   & 0.18  &  \nodata  &  \nodata  &  \nodata \\
S VI 944   & 0.09  &  \nodata  &  \nodata  &  \nodata \\
C III 977\tablenotemark{b}  & 1.15  &  0.14  &  1.55  &  3.97  \\
O VI 1032\tablenotemark{c}  & 3.17  &  0.28  &  16.03\tablenotemark{d} &  10.09\tablenotemark{d} \\
O VI 1038\tablenotemark{c}  & 1.65\tablenotemark{e}  &  0.13  &  \nodata & \nodata \\
\cutinhead{Observed Surface Brightness\tablenotemark{f}
(10$^{-16}$ erg s$^{-1}$ cm$^{-2}$ arcsec$^{-2}$)}
C III 977\tablenotemark{b}  & 1.28  &  1.75  &  0.41  &  7.09  \\
O VI 1032,1038\tablenotemark{c} & 5.36  &  5.13  &  4.28  &  18.02 \\
\enddata
\tablenotetext{a}{Data from Raymond et al.~1997.}
\tablenotetext{b}{Total of both kinematic components observed by FUSE.
55\% is from the component at $-50$~km~s$^{-1}$ and the rest from the
component at $+100$~km~s$^{-1}$.}
\tablenotetext{c}{O~VI is seen only in the $-50$~km~s$^{-1}$ component; see text
and Figure~\protect\ref{figo6c3}.}
\tablenotetext{d}{Total of both lines in the doublet, which was unresolved
in the HUT spectra.}
\tablenotetext{e}{Corrected for C~II $\lambda$1037.02 emission; see text.}
\tablenotetext{f}{Not corrected for foreground extinction.}
\end{deluxetable}


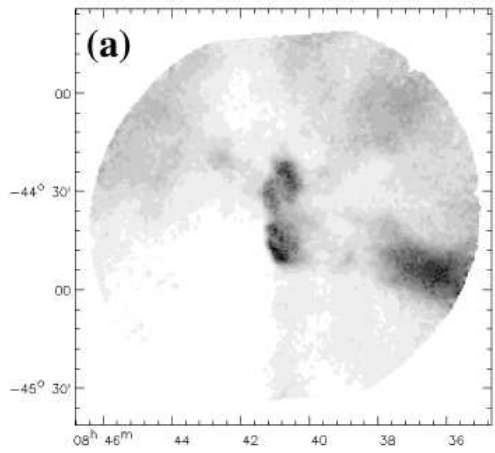
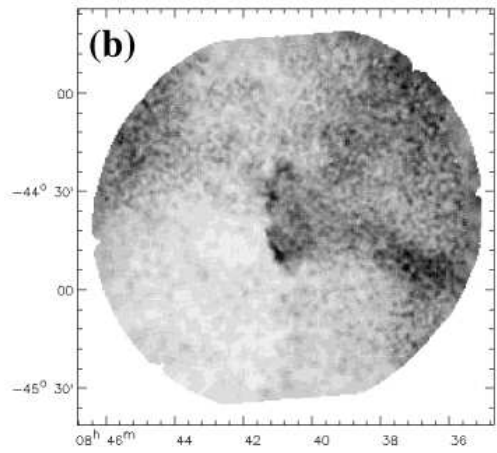
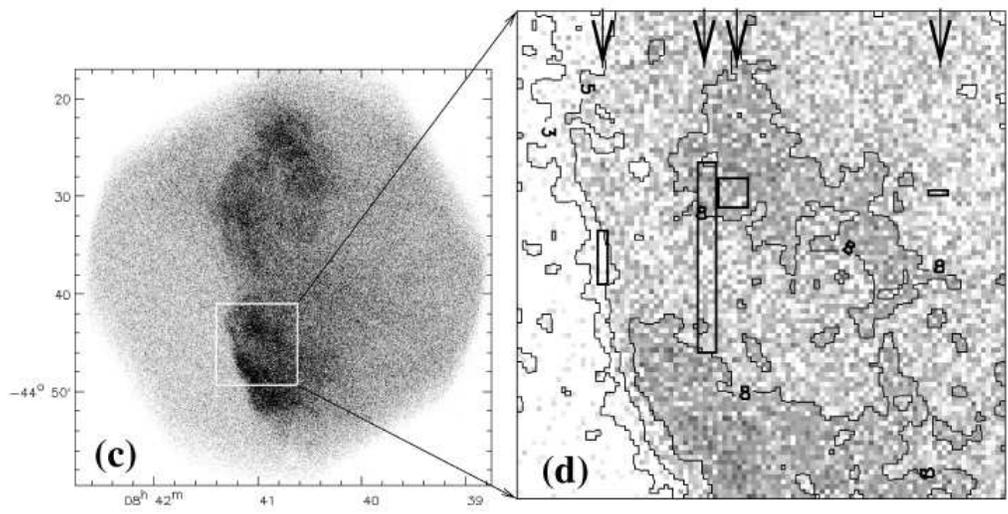
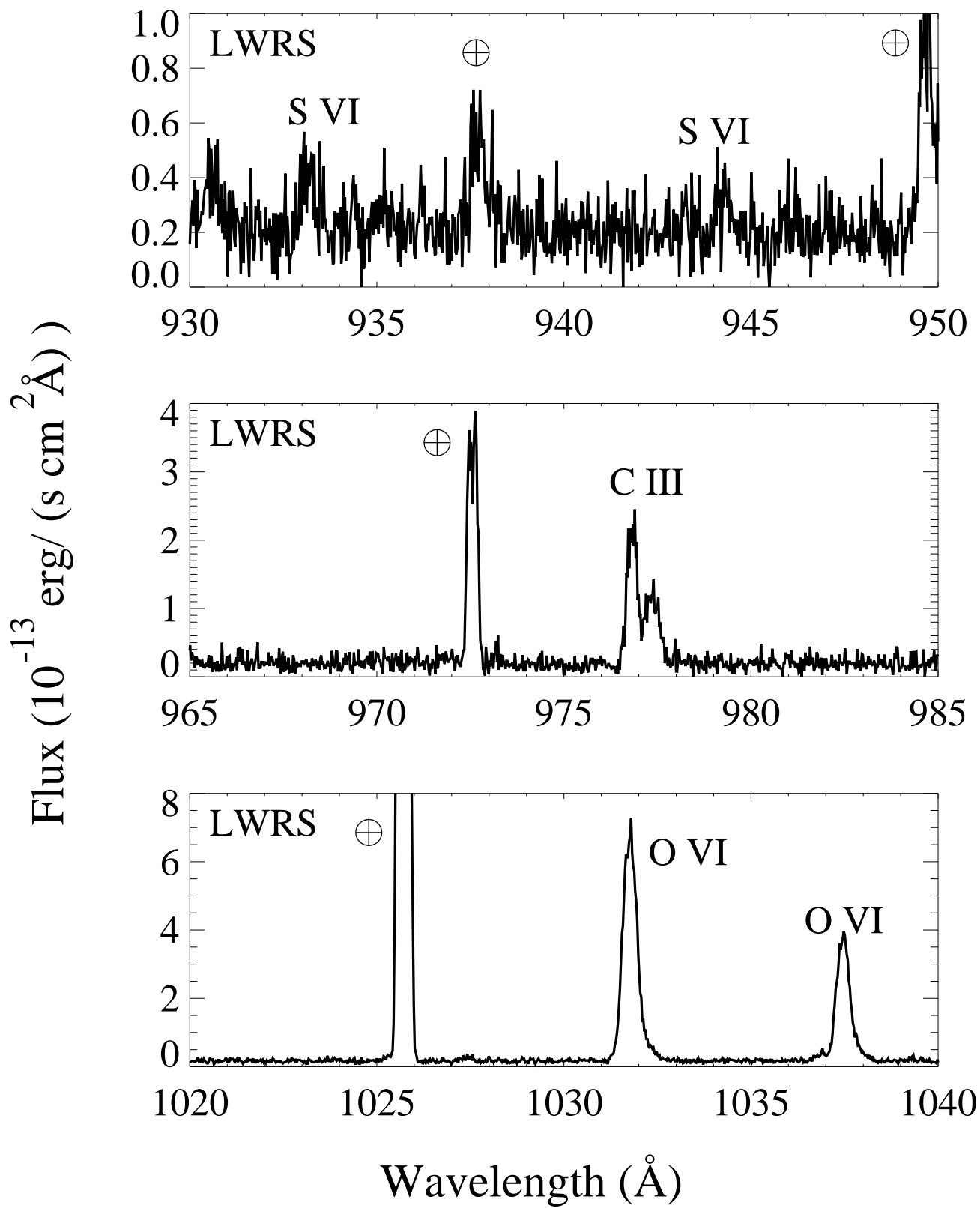
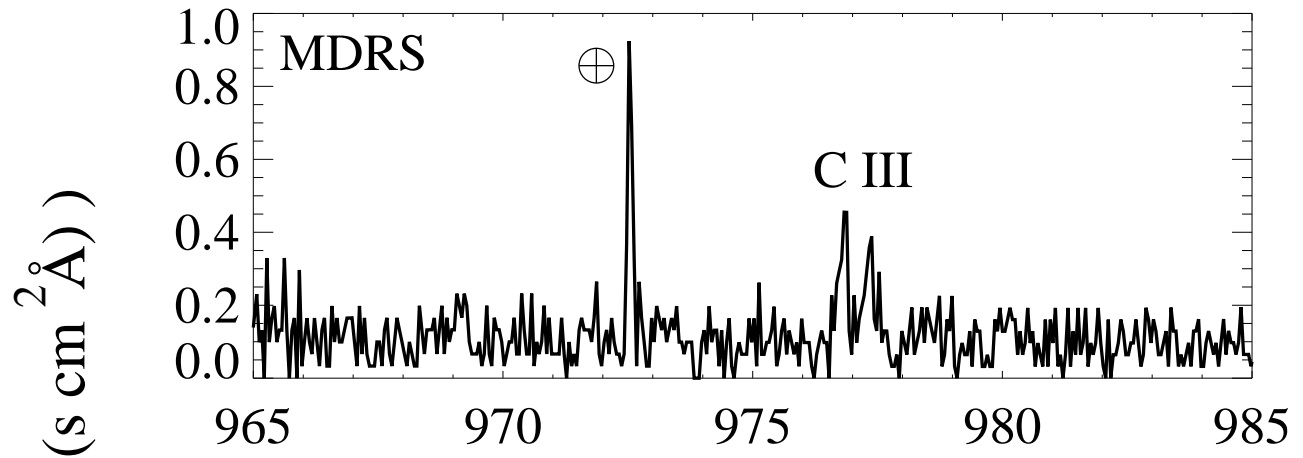
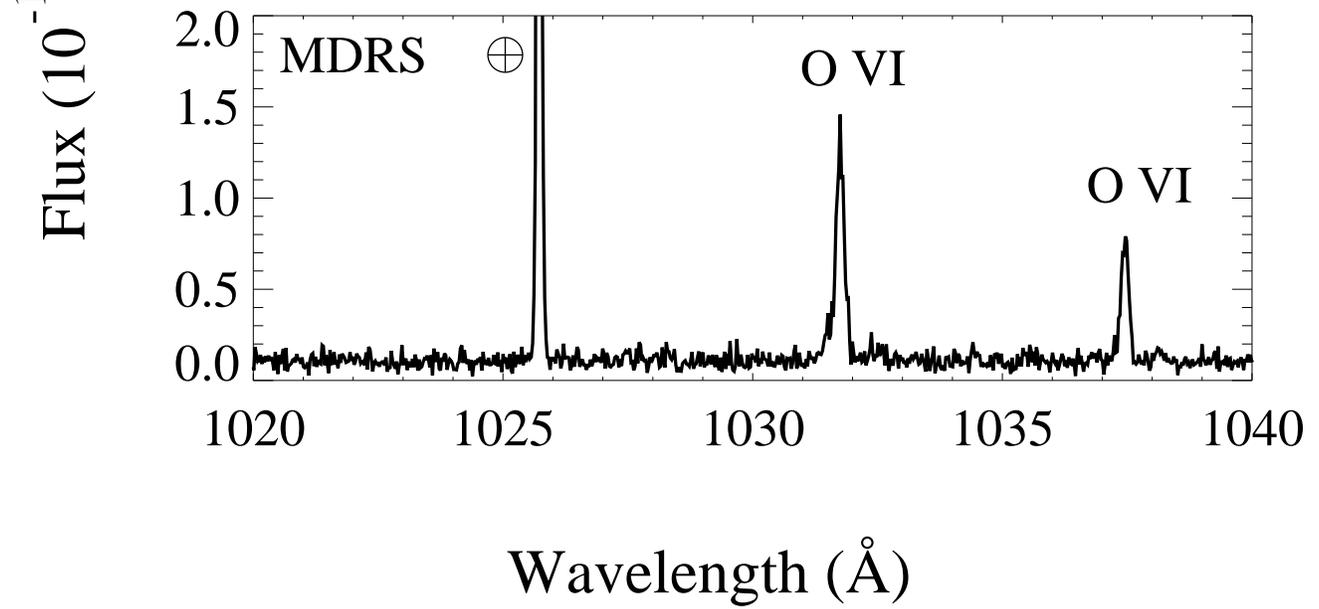
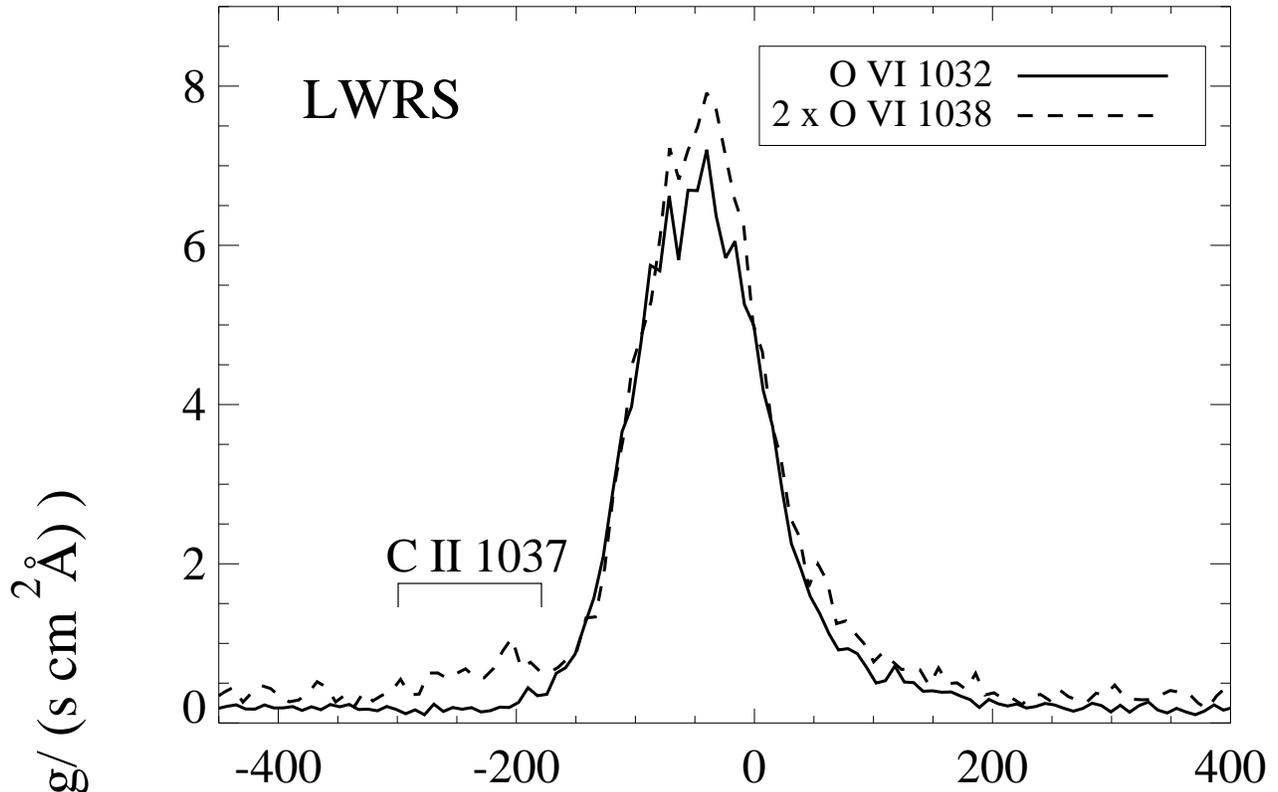
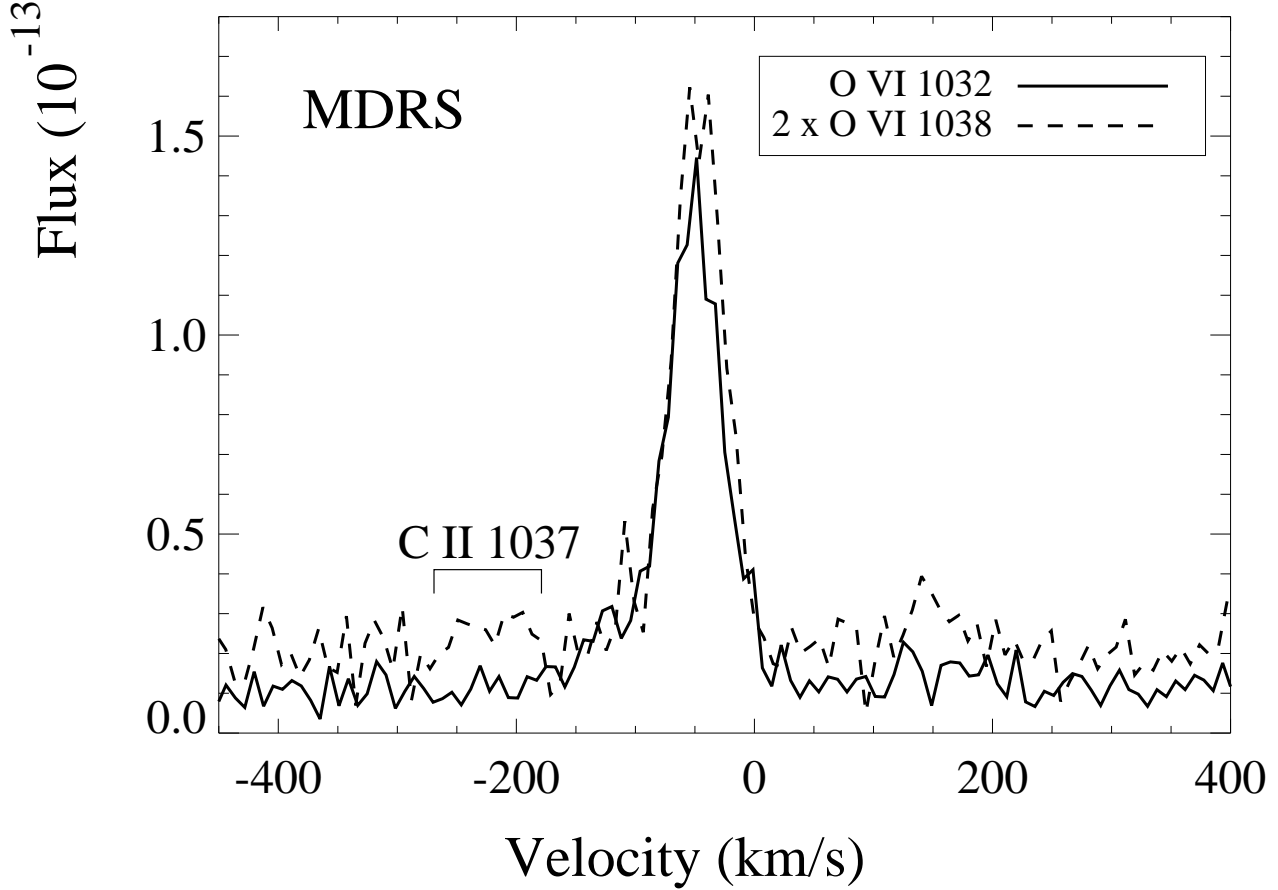
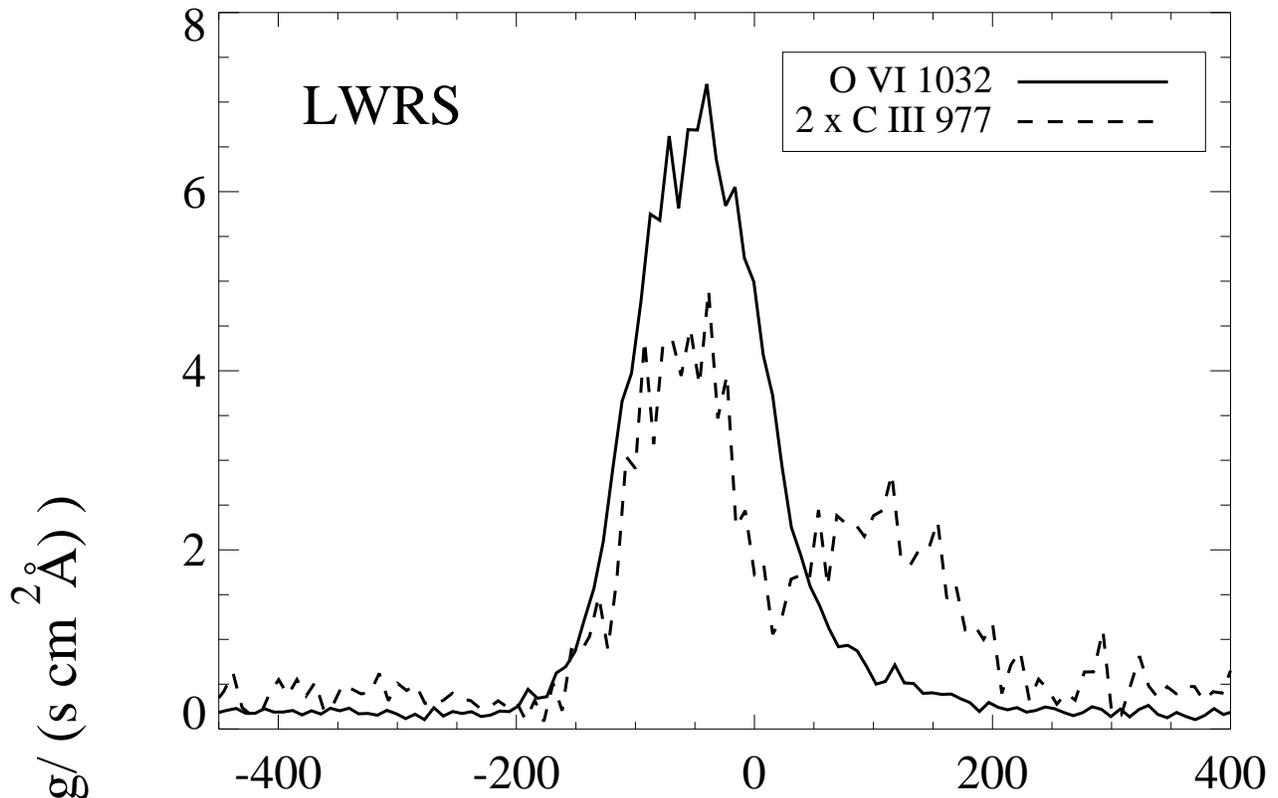
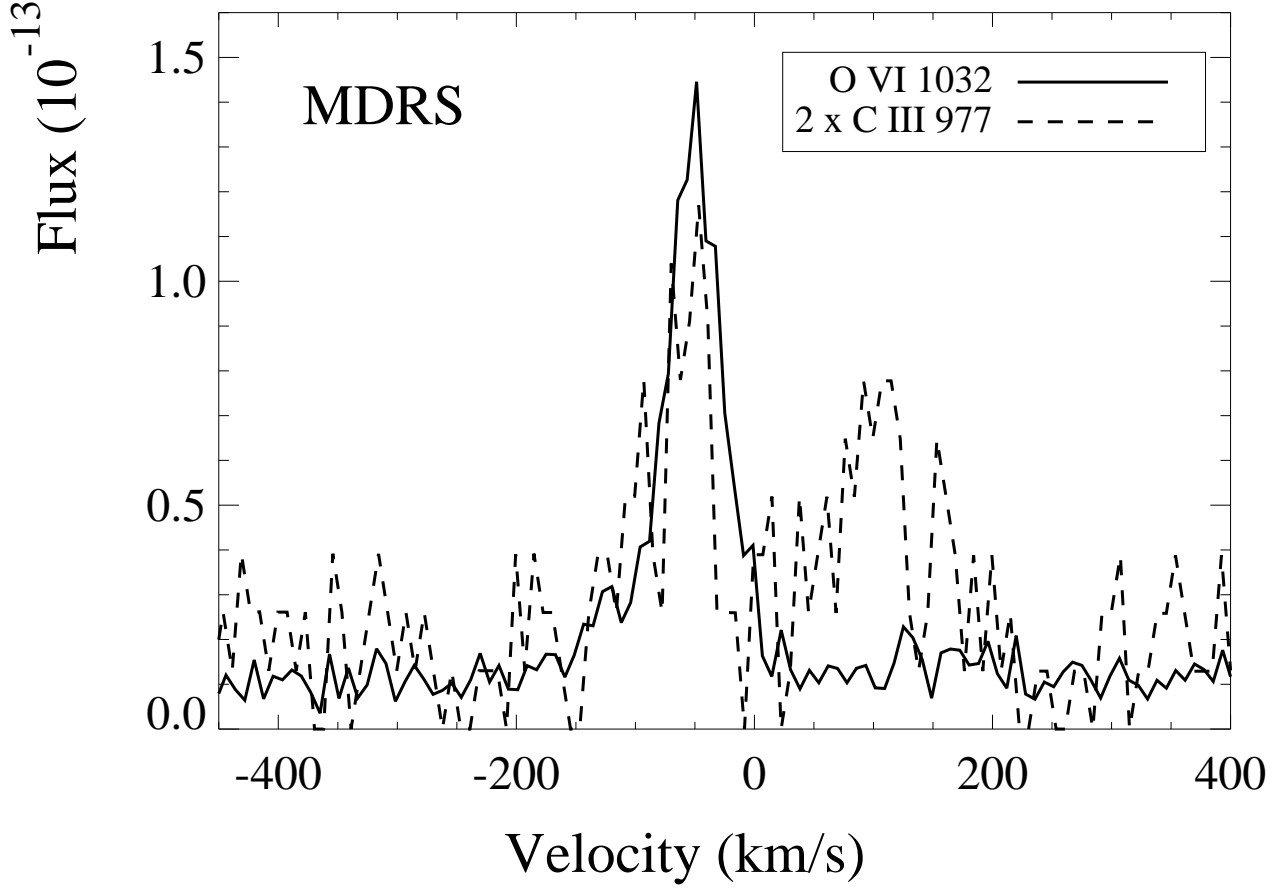
\end{document}